\documentclass[english,aps,preprint,superscriptaddress]{revtex4}
\usepackage[]{fontenc}
\usepackage[latin9]{inputenc}
\usepackage{amsmath}
\usepackage{esint}
\usepackage{epsfig}
\usepackage{graphicx}
\usepackage{bm}
\usepackage{rotating}
\makeatletter
\@ifundefined{textcolor}{}
{%
 \definecolor{BLACK}{gray}{0}
 \definecolor{WHITE}{gray}{1}
 \definecolor{RED}{rgb}{1,0,0}
 \definecolor{GREEN}{rgb}{0,1,0}
 \definecolor{BLUE}{rgb}{0,0,1}
 \definecolor{CYAN}{cmyk}{1,0,0,0}
 \definecolor{MAGENTA}{cmyk}{0,1,0,0}
 \definecolor{YELLOW}{cmyk}{0,0,1,0}
 }

\makeatletter

\newcommand{\be}{\begin{equation}}\newcommand{\ee}{\end{equation}}\newcommand{\ba}{\begin{align}}\newcommand{\ea}{\end{align}}\def\bea{\begin{eqnarray}}\def\eea{\end{eqnarray}}

\usepackage{slashed}

\makeatother

\usepackage{babel}

\begin{document}

\title{Majorana Neutrino Mass Matrices with a Texture Zero and a Cofactor Zero under Current Experimental Texts}

\author{Weijian Wang}

\affiliation{Department of Physics, North China Electric Power
University, Baoding 071003, P. R. China} \affiliation{Zhejiang
Institute of Modern Physics, Zhejiang University, Hangzhou 310027,
P.R. China}

\email{wjnwang96@gmail.com}

\author{Dong-Jiang Zhang}

\affiliation{Department of Physics, Zhejiang University, Hangzhou,
P.R. China}

\begin{abstract}
The Majorana neutrino mass textures with a texture zero and a
vanishing cofactor are reconsidered in the light of current
experimental results. A numerical and systematic analysis is carried
out for all viable patterns. In particular, we focus on the
phenomenological implication of correlations between three mixing
angle (especially for $\theta_{23}$), Dirac CP-violating phase
$\delta$, the effective Majorana neutrino mass $m_{ee}$. We
demonstrated that the correlations between these variables play an
important role in the model selection and can be measured in future
long-baseline oscillation and neutrinoless double beta decay. Among
the six viable patterns, it is the type-III with normal hierarchy
and type-VI with inverted hierarchy that have the parameter space
where the atmospheric neutrino mixing angle $\theta_{23}$ is less
then maximal and the Dirac CP-violating phase covers its best-fit
value.

\vspace{1em}


\end{abstract}
\maketitle By this time, various neutrino oscillation experiments
have provided us with convincing evidences for massive neutrinos and
leptonic flavor mixing with high degree of
accuracy\cite{neu1,neu2,neu3}. The large reactor angle
$\theta_{13}$($\approx 9^{\circ}$)opens the door for us to explore
the leptonic Dirac-CP violation and the mass hierarchy in the future
long-baseline oscillation experiments. The absolute neutrino mass
scale are strongly constrained by the cosmology
observation\cite{WMAP, Planck} and neutrinoless double-beta decay
($0\nu\beta\beta$) experiments (for a review, see \cite{NDD}). On
the other hand, it is still the main theoretical challenge to
understand dynamic origin behind the observed bi-large structure of
leptonic flavor mixing and the mass hierarchy spectrum. Although a
full theory is still missing, several flavor symmetries have been
proposed within the seesaw mechanism to reduce the number of free
parameters in Yukawa sector and reveal the phenomenologically
acceptable mixing pattern. These ideas include texture
zeros\cite{zero}, hybrid textures\cite{hybrid, hybrid2}, zero
trace\cite{sum}, zero determinant\cite{det}, vanishing
minors\cite{minor,minor2}, two traceless submatrices\cite{tra},
equal elements or cofactors\cite{co}, hybrid $M_{\nu}^{-1}$
textures\cite{hyco}. Among these models, the textures with zero
elements or zero minors are particularly appealed, which is not only
because the textures can be naturally realized by introducing proper
flavor symmetry, but also they are stable against the one-loop
quantum corrections as the running of RGEs from seesaw scale
$\Lambda$ to the electroweak scale $\mu\simeq M_{Z}$.

In this work, we reconsider the Majorana neutrino mass textures with
one texture and one vanishing minor, which have been studied in
Ref.\cite{minor2}. These models can be realized by $Z_{12}\times
Z_{2}$ flavor symmetry. There are six types of textures compatible
with the neutrino oscillation data. Our aim is to perform a more
updated and complete analysis on these texture based on the current
experimental data including the large reactor angle and the new
cosmological bound on the sum of neutrino mass. In particular, we
want to analyze the correlations among the Dirac CP-violating phase
$\delta$, the three mixing angle, the Jarlskog invariant $J_{CP}$,
the effective neutrino mass $m_{ee}$.

As is pointed out in Ref.\cite{minor2}, the six textures compatible
with the neutrino oscillation data are given by
\begin{equation}\begin{split}
(M_{\nu})^{I}=\left(\begin{array}{ccc}
\bigtriangleup&0&\times\\
0&\times&\times\\
\times&\times&\times
\end{array}\right)\quad\quad
(M_{\nu})^{II}=\left(\begin{array}{ccc}
\bigtriangleup&\times&0\\
\times&\times&\times\\
0&\times&\times
\end{array}\right)\quad\quad
(M_{\nu})^{III}=\left(\begin{array}{ccc}
\times&0&\times\\
0&\bigtriangleup&\times\\
\times&\times&\times
\end{array}\right)\\
(M_{\nu})^{IV}=\left(\begin{array}{ccc}
\times&\times&0\\
\times&\times&\times\\
0&\times&\bigtriangleup
\end{array}\right)\quad\quad
(M_{\nu})^{V}=\left(\begin{array}{ccc}
\times&\bigtriangleup&\times\\
\bigtriangleup&0&\times\\
0&\times&\times
\end{array}\right)\quad\quad
(M_{\nu})^{VI}=\left(\begin{array}{ccc}
\times&\times&\bigtriangleup\\
\times&\times&\times\\
\bigtriangleup&\times&0
\end{array}\right)
\end{split}\label{matrx}\end{equation}
where the "0" stands for the zero element and the "$\bigtriangleup$"
stands for the zero cofactor.

 In the basis
where the charged mass matrix is diagonal, the neutrino mass texture
$M_{\nu}$ under flavor basis is given by
\begin{equation}
M_{\nu}=VM_{diag}V^{T}\end{equation} where $M_{diag}$ is the
diagonal matrix of neutrino mass eigenvalues $M_{diag}=$diag$(m_{1},
m_{2}, m_{3})$. The Pontecorvo-Maki-Nakagawa-Sakata
matrix\cite{PMNS} $V=U\cdot P$ can be parameterized as
\begin{equation}
V=UP=\left(\begin{array}{ccc}
c_{12}c_{13}&c_{13}s_{12}&s_{13}e^{-i\delta}\\
-s_{12}c_{23}-c_{12}s_{13}s_{23}e^{i\delta}&c_{12}c_{23}-s_{12}s_{13}s_{23}e^{i\delta}&c_{13}s_{23}\\
s_{23}s_{12}-c_{12}c_{23}s_{13}e^{i\delta}&-c_{12}s_{23}-c_{23}s_{12}s_{13}e^{i\delta}&c_{13}c_{23}
\end{array}\right)\left(\begin{array}{ccc}
1&0&0\\
0&e^{i\alpha}&0\\
0&0&e^{i(\beta+\delta)}
\end{array}\right)
\label{3}\end{equation} where the abbreviation
$s_{ij}=\sin\theta_{ij}$ and $c_{ij}=\cos\theta_{ij}$ is used. In
neutrino oscillation experiments, CP violation effect is usually
reflected by the Jarlskog rephasing invariant quantity\cite{Jas}
defined as
\begin{equation}
J_{CP}=s_{12}s_{23}s_{13}c_{12}c_{23}c_{13}^{2}\sin\delta
\end{equation}

Following the same step in Ref.\cite{minor2}, the texture zero and
the zero minor in $M_{\nu}$ gives the mass ratio
$(\frac{m_{1}}{m_{2}}, \frac{m_{1}}{m_{3}})$ and Majorana
CP-violating phases $(\alpha, \beta)$ in terms of the $(\theta_{12},
\theta_{23},\theta_{13}, \delta)$.i.e
\begin{equation}
\frac{m_{1}}{m_{2}}e^{-2i\beta}=-\frac{K_{1}L_{1}-K_{2}L_{2}+K_{3}L_{3}\pm
(K_{1}^{2}L_{1}^{2}+(K_{2}L_{2}-K_{3}L_{3})^{2}-2K_{1}L_{1}(K_{2}L_{2}+K_{3}L_{3}))^{\frac{1}{2}}}{2K_{1}L_{3}}e^{2i\delta}
\label{ra1}\end{equation} and
\begin{equation}
\frac{m_{1}}{m_{3}}e^{-2i\beta}=\frac{-K_{1}L_{1}-K_{2}L_{2}+K_{3}L_{3}\pm
(K_{1}^{2}L_{1}^{2}+(K_{2}L_{2}-K_{3}L_{3})^{2}-2K_{1}L_{1}(K_{2}L_{2}+K_{3}L_{3}))^{\frac{1}{2}}}{2K_{1}L_{2}}
\label{ra2}\end{equation} where $K_{1}=U_{x1}U_{y1},
K_{2}=U_{x2}U_{y2},K_{3}=U_{x3}U_{y3}$ and
\begin{equation}
L_{i}=(U_{pj}U_{qj}U_{rk}U_{sk}-U_{tj}U_{uj}U_{vk}U_{wk})+(j\leftrightarrow
k)
\end{equation}
with $(i,j,k)$ a cyclic permutation of (1,2,3). With the help
of\eqref{ra1} and \eqref{ra2}, the magnitudes of mass radios are
\begin{equation}
\rho=\Big|\frac{m_{1}}{m_{3}}e^{-2i\beta}\Big|,\quad\quad
\sigma=\Big|\frac{m_{1}}{m_{2}}e^{-2i\alpha}\Big|
\label{tt1}\end{equation} as well as the two Majorana CP-violating
phases
\begin{equation}
\alpha=-\frac{1}{2}arg\Big(\frac{m_{1}}{m_{2}}e^{-2i\alpha}\Big),\quad\quad
\beta=-\frac{1}{2}arg\Big(\frac{m_{1}}{m_{3}}e^{-2i\beta}\Big)
\label{tt2}\end{equation} The neutrino mass ratios $\rho$ and
$\sigma$ are related to the ratios of two neutrino mass-squared
ratios obtained from the solar and atmosphere oscillation
experiments as
\begin{equation}
R_{\nu}\equiv\frac{\delta m^{2}}{|\Delta
m^{2}|}=\frac{2\rho^{2}(1-\sigma^{2})}{|2\sigma^{2}-\rho^{2}-\rho^{2}\sigma^{2}|}
\label{rv}\end{equation} where $\delta m^{2}\equiv
m_{2}^{2}-m_{1}^{2}$ and $\Delta m^{2}\equiv m_{3}^{2}-m_{1}^{2}$.
For normal neutrino mass hierarchy(NH), the latest global-fit
neutrino oscillation experimental data, at the 1$\sigma$, 2$\sigma$
and 3$\sigma$ confidential level, is list as follows\cite{data}
\begin{equation}\begin{split}
\theta_{12}=33.6^{\circ (+1.2,+2.2,+3.3)}_{(-1.0,-2.0,-3.0)}\\
\theta_{23}=38.4^{\circ (+1.6,+3.6,+14.6)}_{(-1.2,-2.2,-3.2)}\\
\theta_{13}=8.9^{\circ (+0.5,+0.9,+1.3)}_{(-0.4,-0.9,-1.4)}\\
\delta m^{2}=7.54^{(+0.26,+0.46,+0.64)}_{(-0.22,-0.39,-0.55)}\times
10^{-5} eV^{2}\\
\bigtriangleup
m^{2}=2.43^{(+0.06,+0.12,+0.19)}_{(-0.10,-0.16,-0.24)}\times 10^{-3}
eV^{2}
\end{split}\end{equation}
For the inverted neutrino mass hierarchy(NH), the differences
compared with the NH are so slight that we shall use the same values
given above. It is noted that the global analysis tends to give a
$\theta_{23}$ less than $45^{\circ}$ at 2$\sigma$ and 1$\sigma$
level. The Majorana nature of neutrino can be determined if any
signal of neutrinoless double decay is observed, implying the
violation of leptonic number violation. The decay ratio is related
to the effective of neutrino $m_{ee}$, which is written as
\begin{equation}
m_{ee}=|m_{1}c_{12}^{2}c_{13}^{2}+m_{2}s_{12}^{2}c_{13}^{2}e^{2i\alpha}+m_{3}s_{13}^{2}e^{2i\beta}|
\end{equation}
 Although a $3\sigma$ result of $m_{ee}=(0.11-0.56)$ eV is
reported by the Heidelberg-Moscow Collaboration\cite{HM}, this
result is criticized in Ref \cite{NND2} and shall be checked by the
forthcoming experiment. It is believed that that the next generation
$0\nu\beta\beta$ experiments, with the sensitivity of $ m_{ee}$
being up to 0.01 eV, will open the window to not only the absolute
neutrino mass scale but also the Majorana-type CP violation. Besides
the $0\nu\beta\beta$ experiments, a more severe constraint was set
from the recent cosmology observation. Recently, an upper bound on
the sum of neutrino mass $\sum m_{i}<0.23$ eV is
reported\cite{Planck} by Plank Collaboration combined with the WMAP,
high-resolution CMB and BAO experiments.

In the numerical analysis, We randomly vary the three mixing angles
$(\theta_{12}, \theta_{23}, \theta_{13})$in their $3\sigma$ range.
Up to now, no bound was set on Dirac CP-violating phase $\delta$ at
3 $\sigma$ level, so we vary it randomly in the range of $[0,2\pi]$.
Using Eq. \eqref{rv}, the mass-squared difference ratio $R_{\nu}$ is
determined. Then the input parameters is empirically acceptable when
the $R_{\nu}$ falls inside the the $3\sigma$ range of experimental
data, otherwise they are excluded. Finally, we get the value of
neutrino mass and Majorana CP-violating $\alpha$ and $\beta$ though
Eq.\eqref{tt1}, \eqref{tt2}. Since we have already obtained the
absolute neutrino mass $m_{1,2,3}$ and $(\alpha,\beta)$, the further
constraint from cosmology should be considered. In this work, we set
the upper bound on the sum of neutrino mass $\Sigma m_{i}$ less than
0.23 eV.

\begin{figure}
 \includegraphics[scale = 0.50]{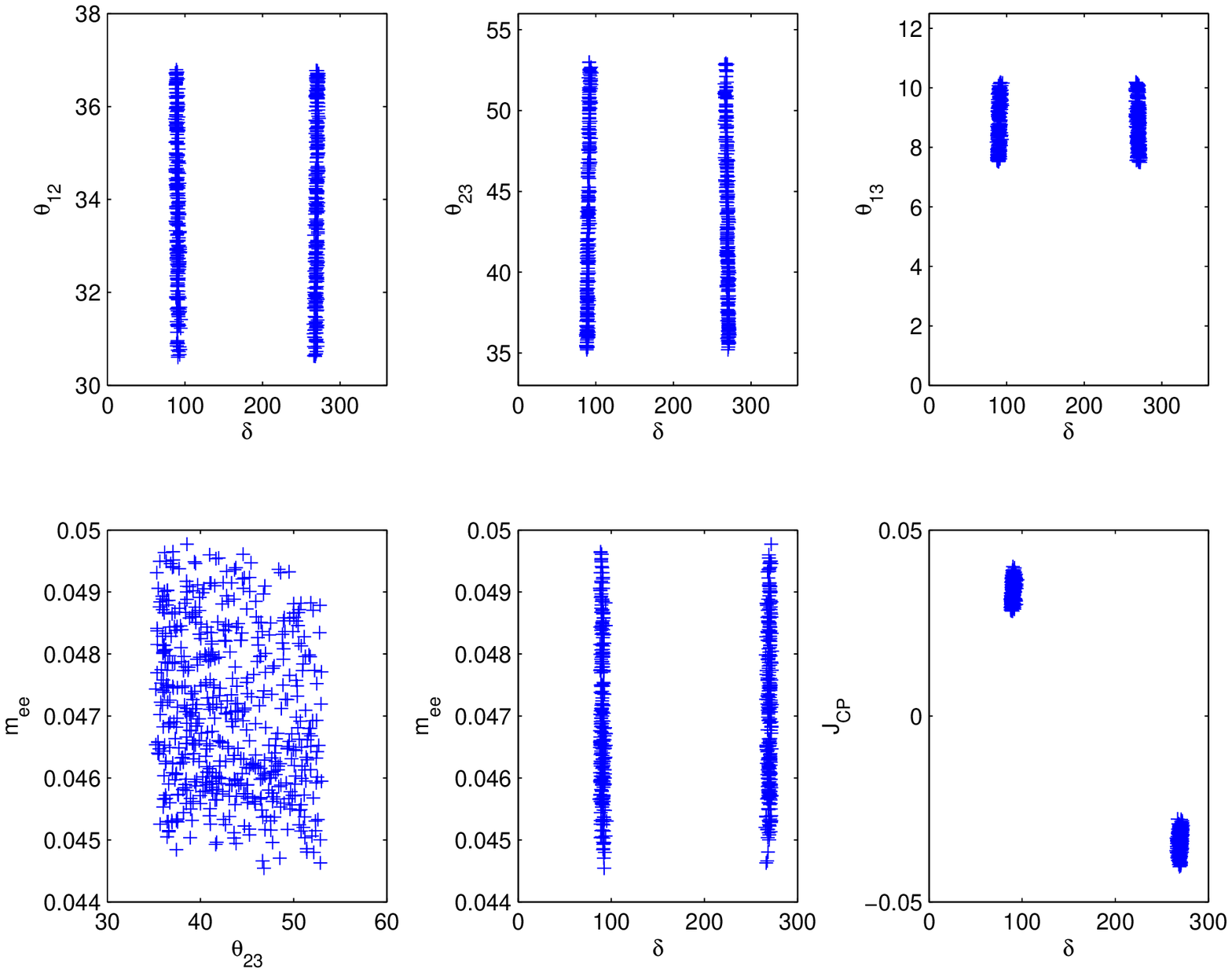}%
\caption {The plots for pattern type-I (IH). } \label{2AIH}
 \end{figure}

\begin{figure}
 \includegraphics[scale = 0.50]{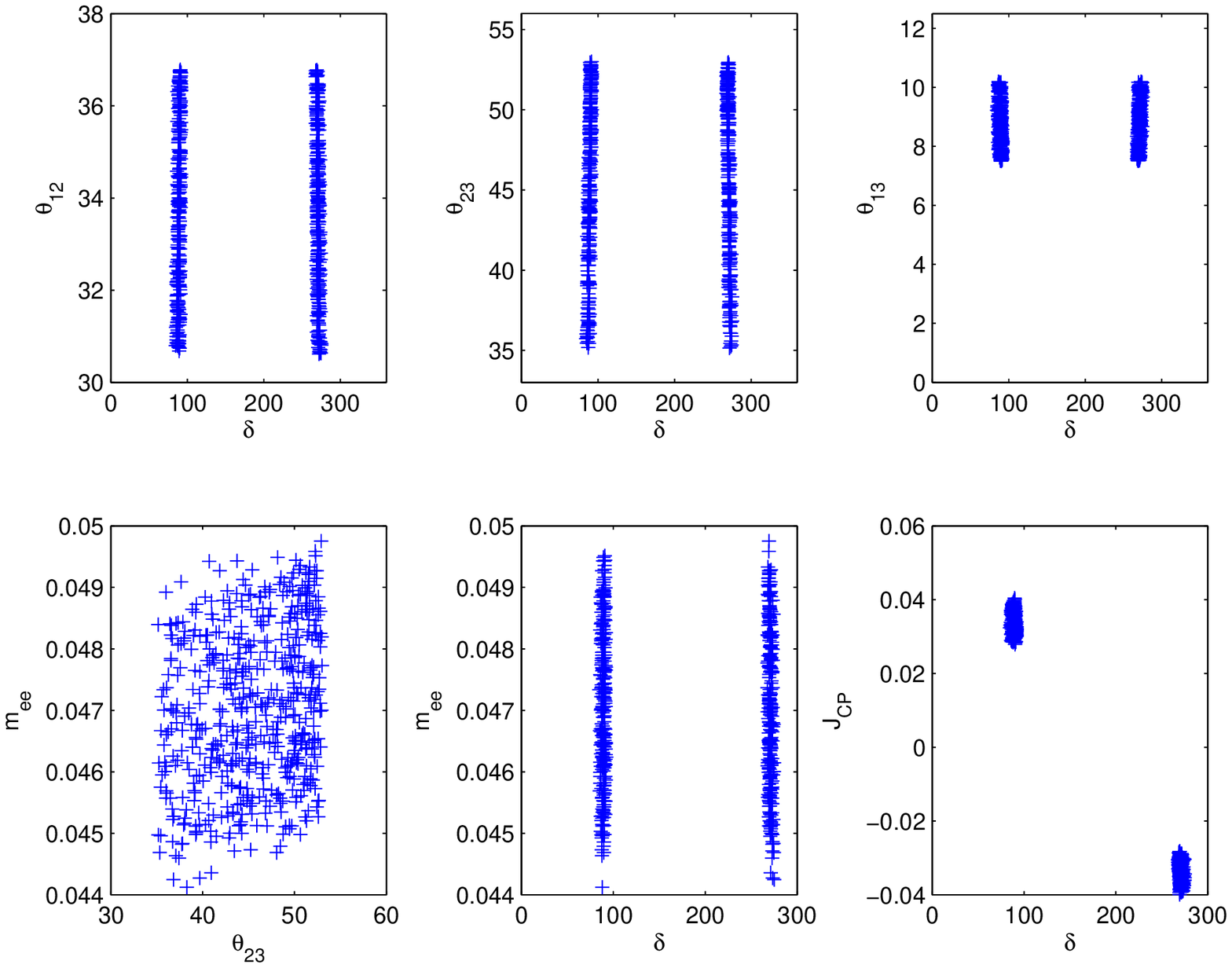}%
\caption {The plots for pattern type-II (IH). } \label{3AIH}
 \end{figure}

\begin{figure}
 \includegraphics[scale = 0.50]{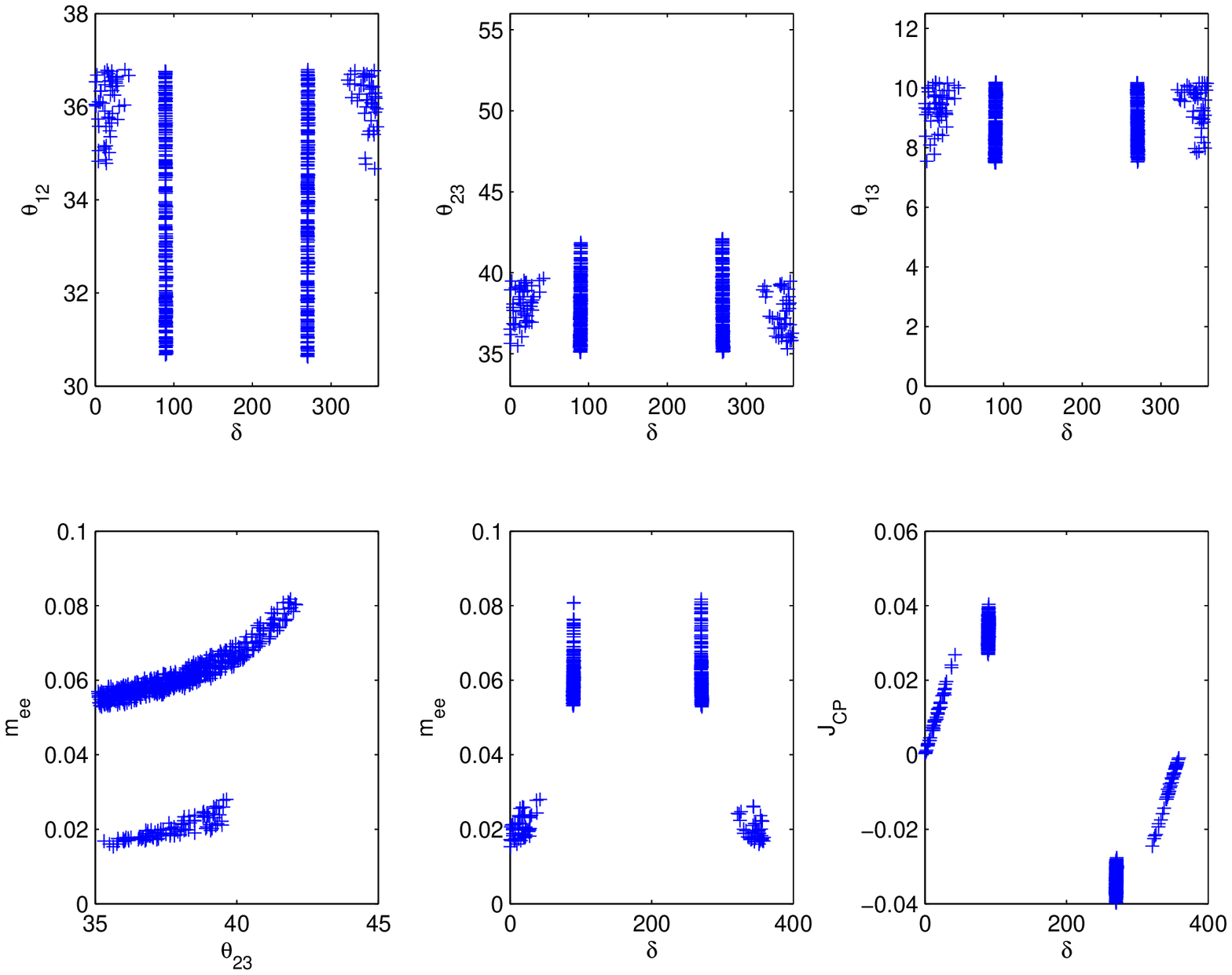}%
\caption {The plots for pattern type-III (IH). } \label{2DIH}
 \end{figure}

\begin{figure}
 \includegraphics[scale = 0.50]{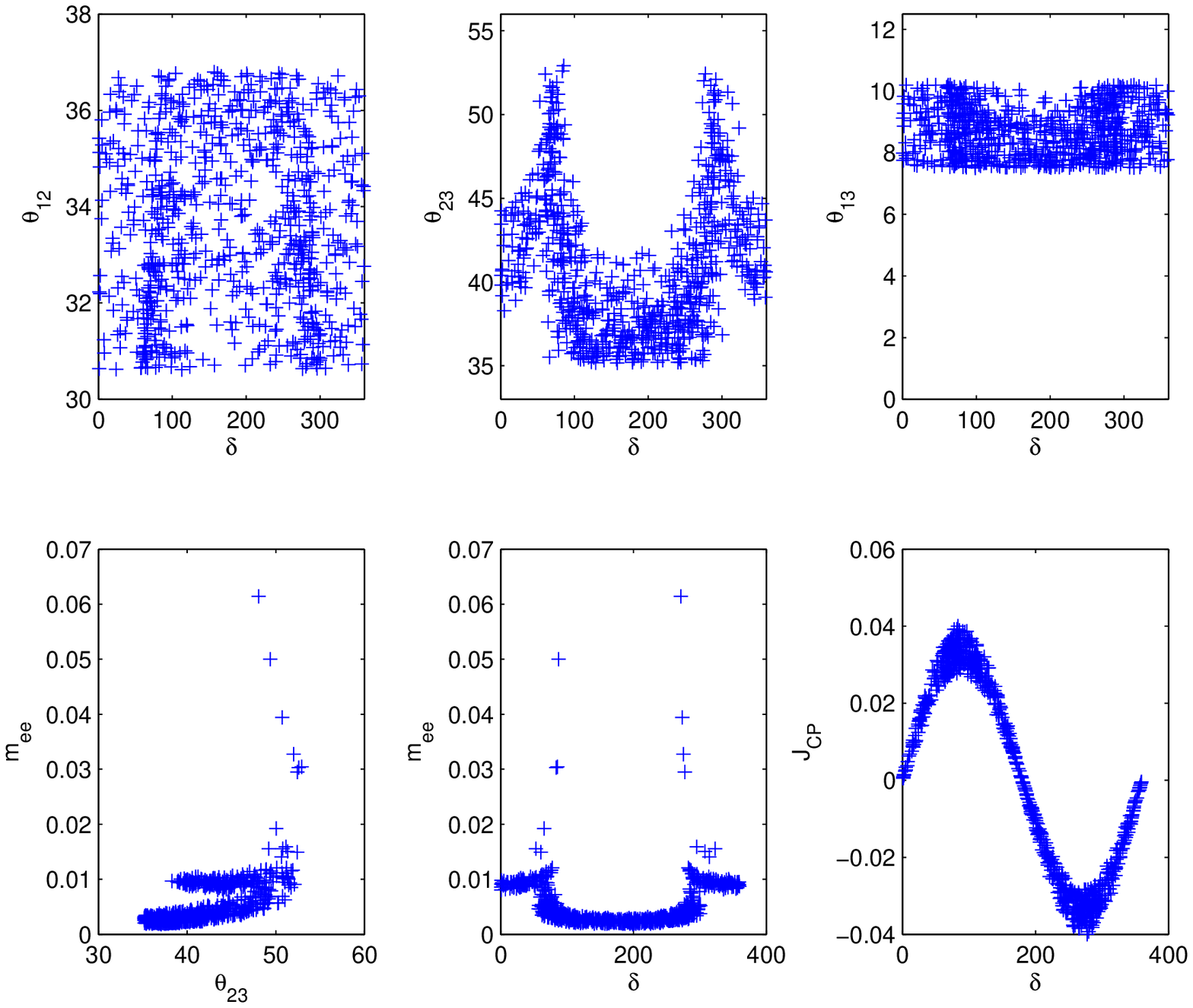}%
\caption {The plots for pattern type-III (NH). } \label{2DNH}
 \end{figure}

\begin{figure}
 \includegraphics[scale = 0.50]{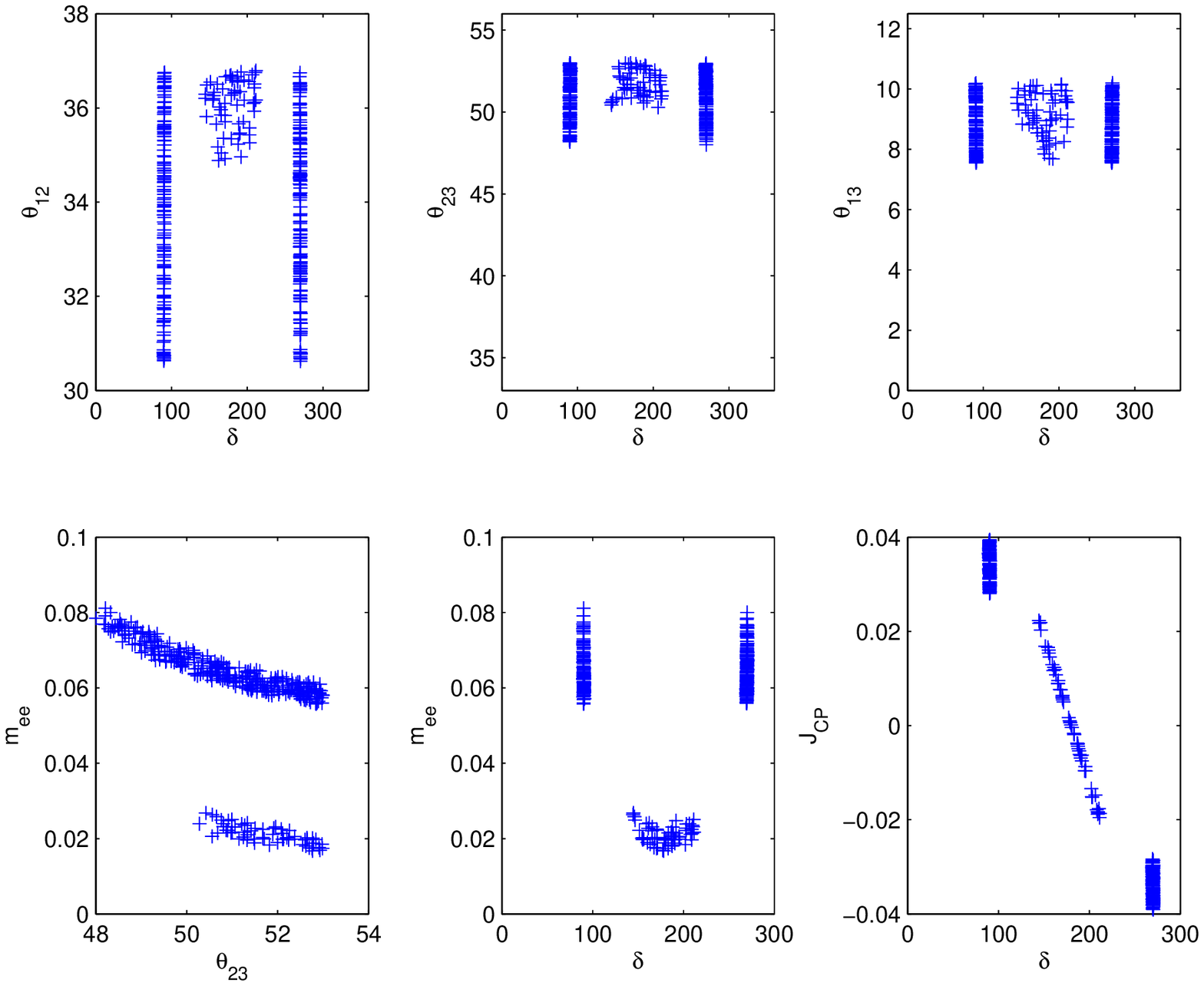}%
\caption {The plots for pattern type-IV (IH). } \label{3FIH}
 \end{figure}

\begin{figure}
 \includegraphics[scale = 0.50]{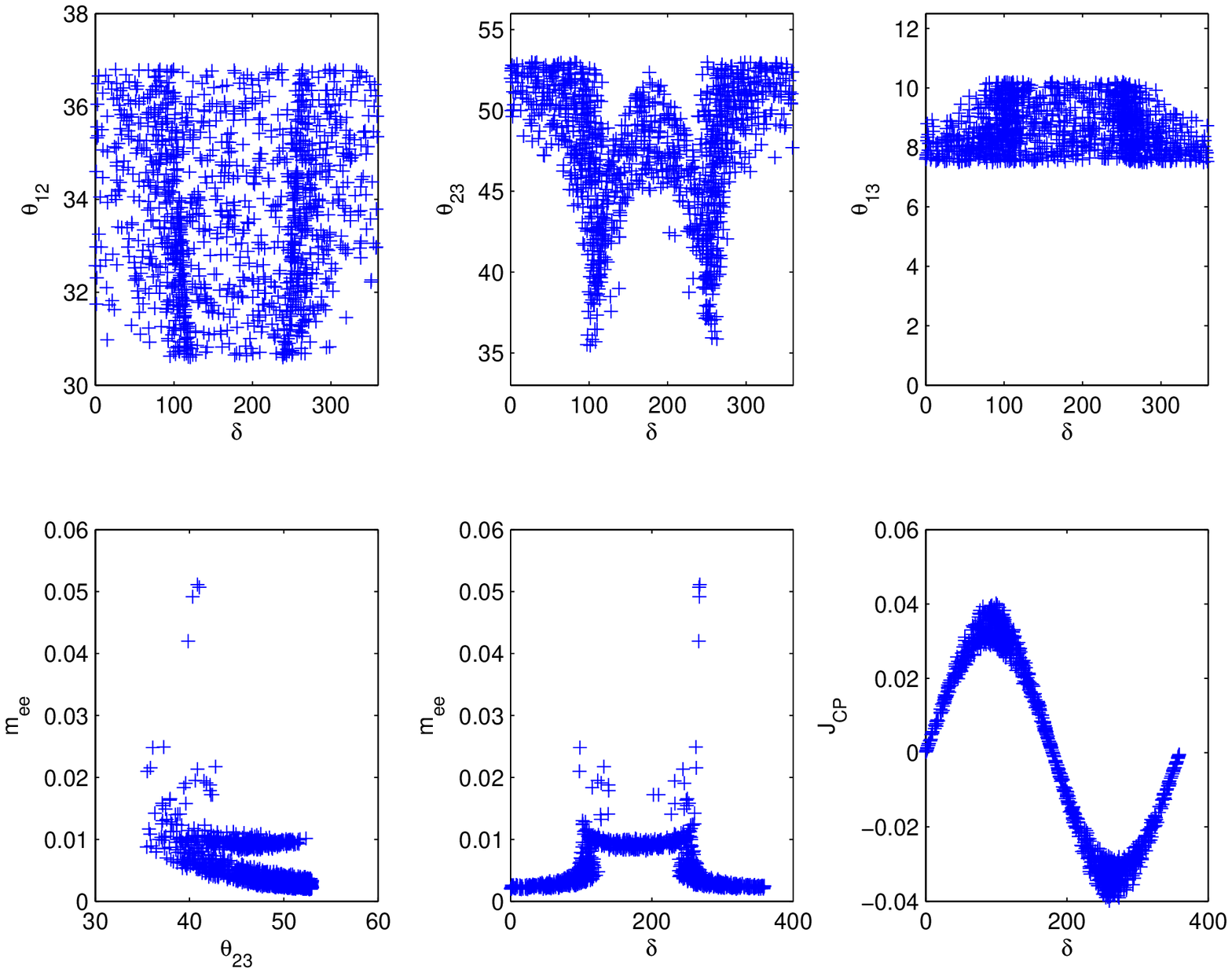}%
\caption {The plots for pattern type-IV (NH). } \label{3FNH}
 \end{figure}

\begin{figure}
 \includegraphics[scale = 0.50]{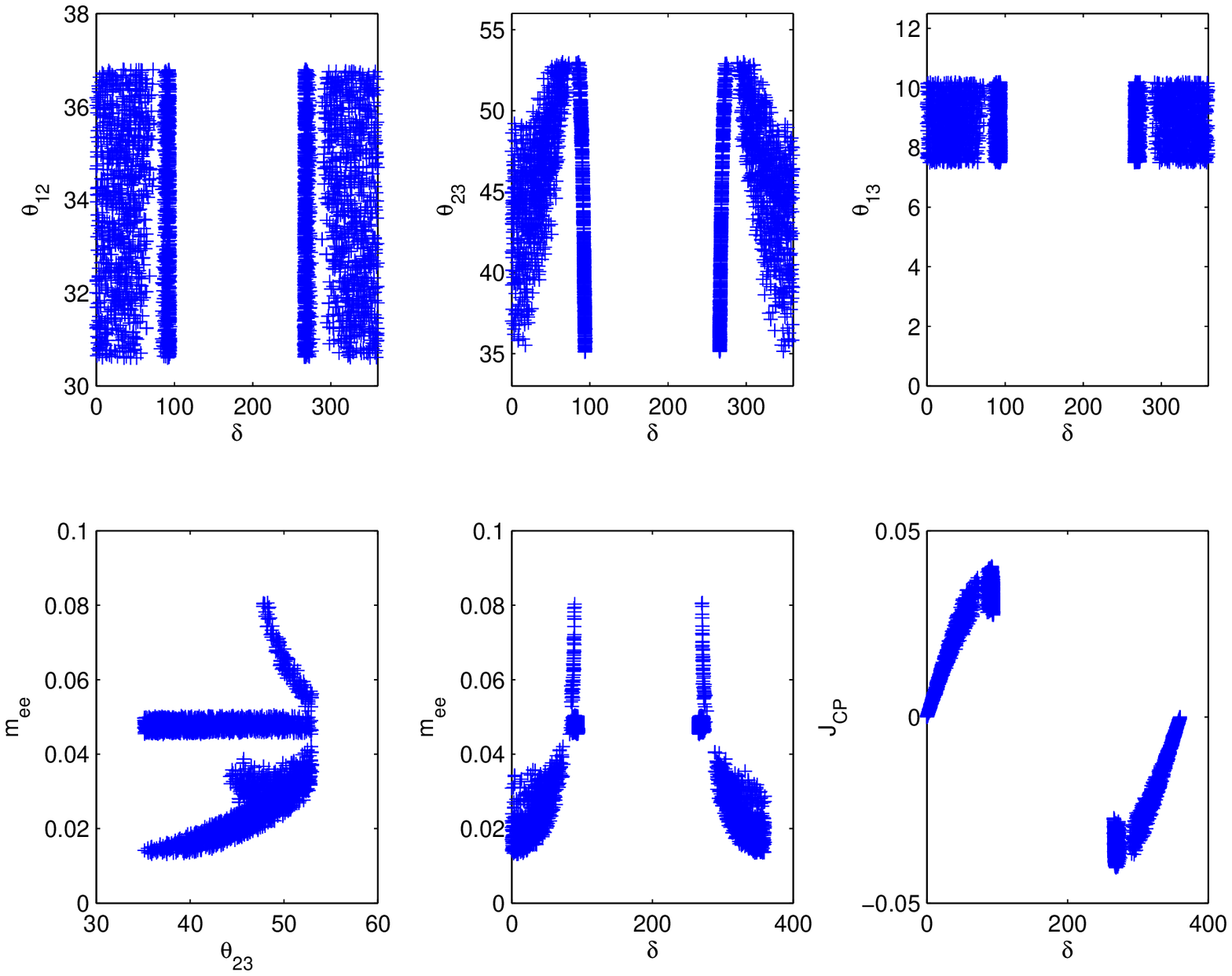}%
\caption {The plots for pattern type-V (IH). } \label{4BIH}
 \end{figure}

\begin{figure}
 \includegraphics[scale = 0.50]{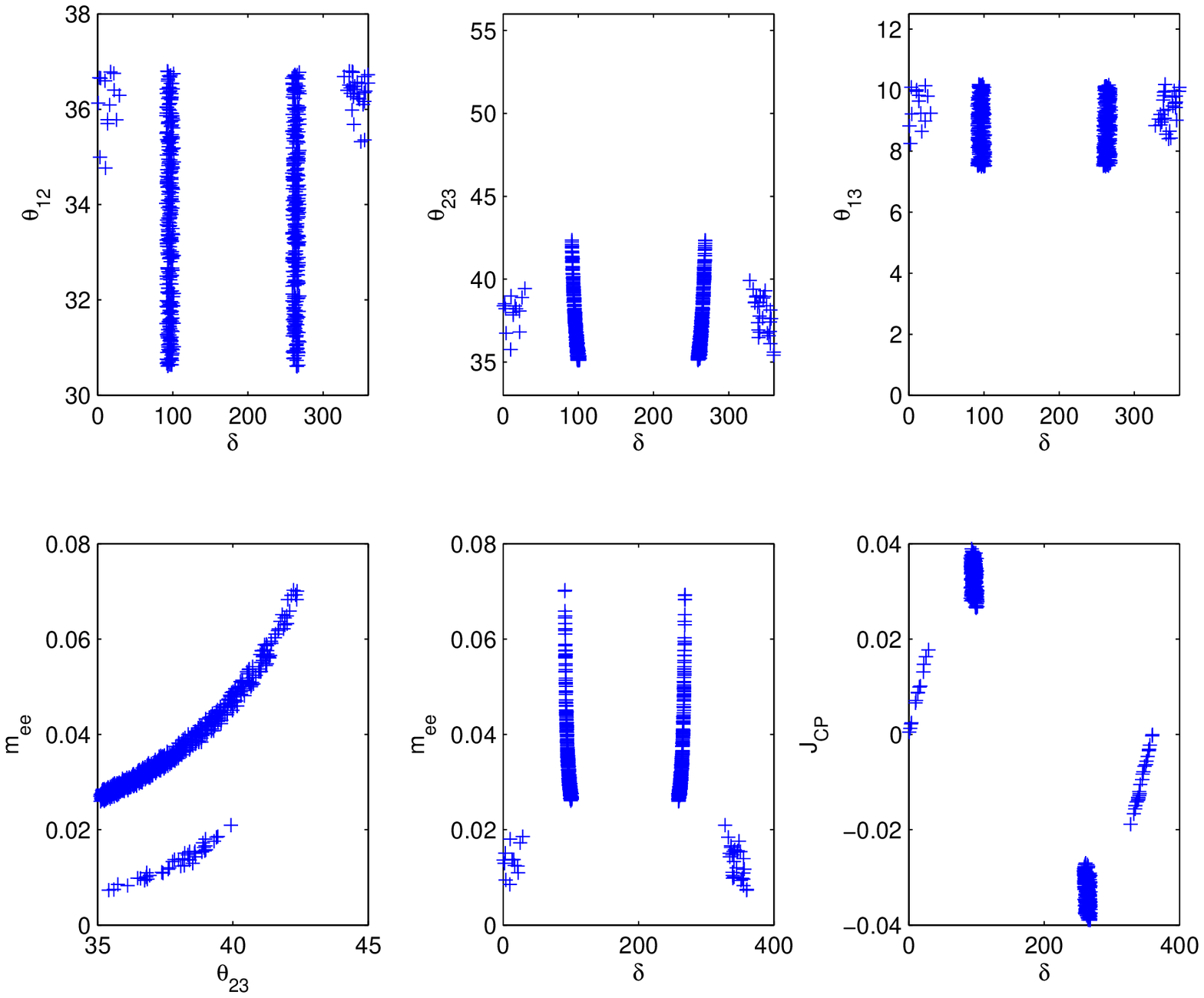}%
\caption {The plots for pattern type-V (NH). } \label{4BNH}
 \end{figure}

\begin{figure}
 \includegraphics[scale = 0.50]{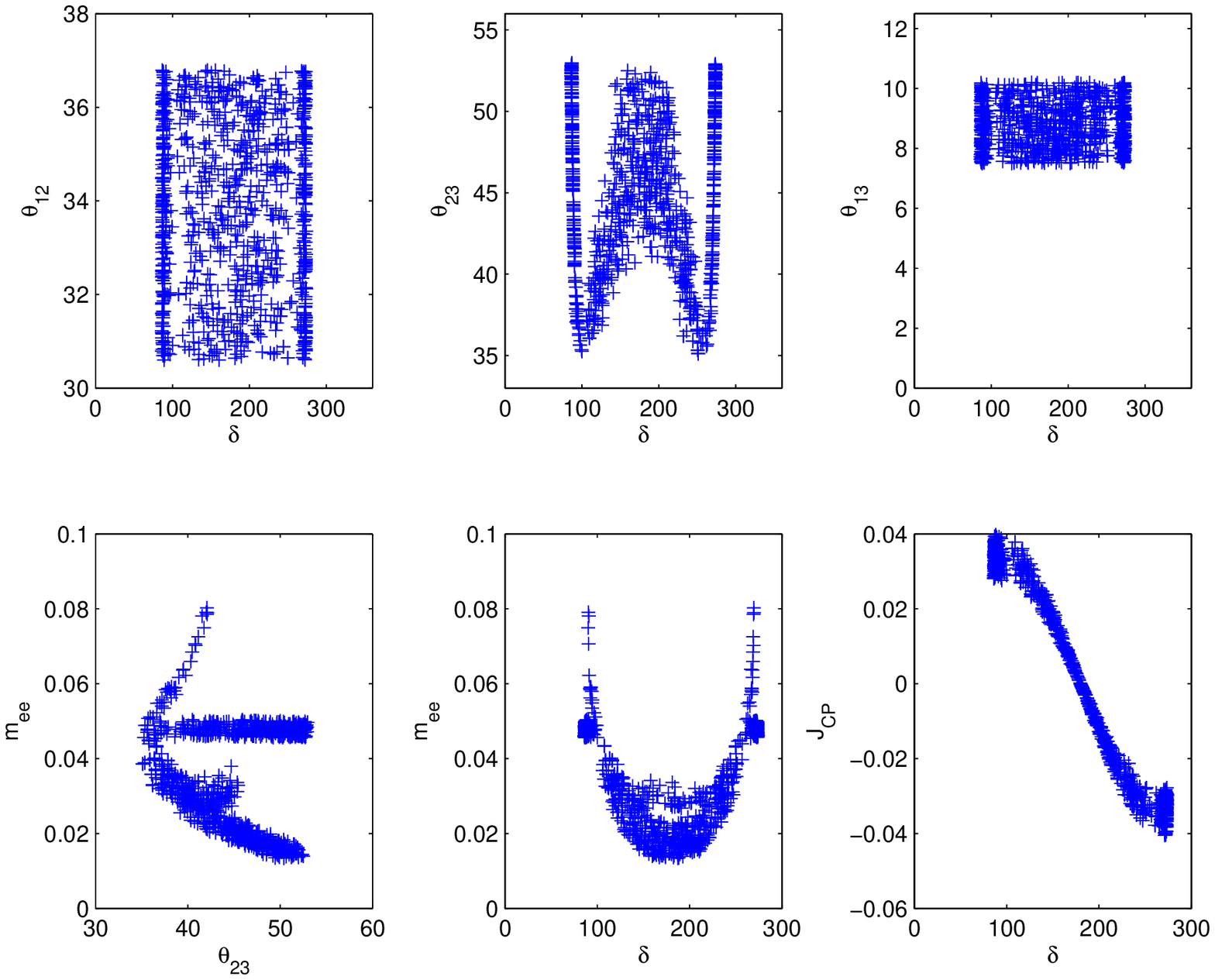}%
\caption {The plots for pattern type-VI (IH). } \label{6CIH}
 \end{figure}

\begin{figure}
 \includegraphics[scale = 0.50]{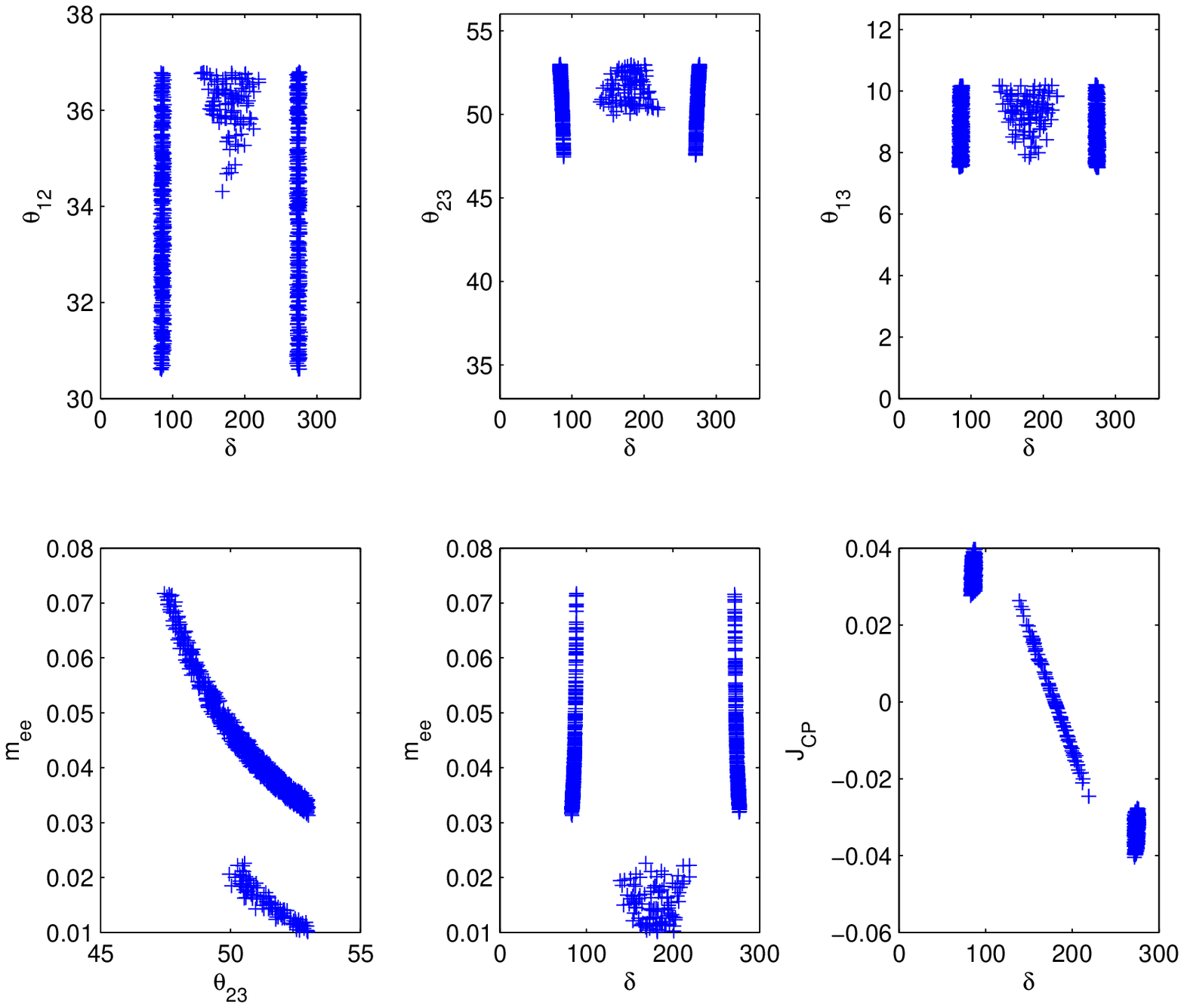}%
\caption {The plots for pattern type-VI (NH). } \label{6CNH}
 \end{figure}

In Fig.\ref{2AIH}-\ref{6CNH}, we demonstrate the correlations for
all six pattern. The main results and the discussion are summarized
as follows:

1) The Type-I and Type-II patterns are phenomenological acceptable
only for inverted mass hierarchy, as mentioned in Ref.\cite{minor2}.
The type-I pattern are related to the type-II pattern by the
$\mu-\tau$ symmetry\cite{mut}, leading to the similar allowed resign
for both patterns. One can see from the Fig.\ref{2AIH} and Fig.
\ref{3AIH} that in the light of a large $\theta_{13}$ the Dirac
CP-violating phase $\delta$ can only acceptable at around
$\pi/2(3\pi/2)$. The strong bound on $\delta$ is unusual and can be
verified or excluded in future experiments. The value of
$\theta_{23}$ and $\theta_{12}$ are fully covered in $3\sigma$ level
range. The $m_{ee}$ for both type-I and type-II pattern lie in the
range of $0.044$eV$<m_{ee}<0.05$eV, which is in the scope of
accuracy of $0\nu \beta\beta$ decay experiment near the future. For
the maximal CP-violating $\delta=\pi/2$ or $\delta=3\pi/2$, the
$|J_{CP}|\geq 3\%$ is achieved, which is accessible to the future
long-baseline neutrino oscillation experiments.

2)The allowed region of type-III patterns are shown in
Fig.\eqref{2DIH} for inverted hierarchy and Fig.\eqref{2DNH} for
normal hierarchy. For the IH pattern, $\theta_{23}$ lies below the
maximality. Interestingly, the $\theta_{12}>34^{\circ}$ and
$\theta_{23}<40^{\circ}$ are satisfied when the Dirac CP-violating
phase $\delta$ falls into the region of $0^{\circ}\sim 50^{\circ}$
($310^{\circ}\sim 360^{\circ}$), leading to $0.015$eV$<m_{ee}<$
0.030eV and $|J_{CP}|<0.03$. However, $\delta$ is also highly
constrained at around $\pi/2$ ($3\pi/2$), where the full $3\sigma$
range of mixing angle are covered and $0.050$eV$<m_{ee}<$ 0.090eV.
The Jarlskog invariant $J_{CP}$ is close to its maximum, i.e.
$|J_{CP}|\geq 3\%$,which is promising to be explored in the
next-generation long-baseline neutrino oscillation experiment. On
the other hand, the Fig.\eqref{2DNH} illustrates more complicated
correlations for the NH pattern. There are unconstraint parameter
space for $\delta$, $\theta_{12}$, $\theta_{13}$ and $J_{CP}$.
However, we get a strongly constrained $\theta_{23}>\pi/4$ if
$\delta$ lie in the range of $120^{\circ}\sim 240^{\circ}$ and
$m_{ee}<0.005$eV, rendering it very challenging to be detected in
future $0\nu\beta\beta$ experiments.

3)We present the scatter plots of type-IV in Figure.\eqref{3FIH} for
IN case and Figure.\eqref{3FNH} for NH case. For the inverted
hierarchy, the Dirac CP-violating phase $\delta$ is limited to two
regions, i.e. $120^{\circ}\sim 240^{\circ}$ and highly constrained
points $\delta=\pi/2$ ($3\pi/2$). The $\theta_{23}$ are all above
maximal and thus phenomenological ruled out at $2\sigma$ level. The
range of $m_{ee}$ is the same to that of type-III. This is not a
coincidence but because the type-IV and and type-III are related by
the $\mu-\tau$ symmetry. Thus the $\theta^{IV}_{23}$ and
$\delta^{IV}$ of type-IV are respectively equal to the
$\frac{\pi}{4}-\theta^{III}_{23}$ and $\pi-\delta^{III}$ of type-III
pattern. The same situation also appears in NH case.

4) The allowed region of type-V pattern are illustrated in
Fig.\eqref{4BIH} and Fig.\eqref{4BNH}. One can see from the figure
that for the IH case, the Dirac CP-violating phase $\delta$ are
phenomenological acceptable in the range of $0^{\circ}\sim
80^{\circ}$( $280^{\circ}\sim 360^{\circ}$)and $90^{\circ}\sim
95^{\circ}$($265^{\circ}\sim 270^{\circ}$).  Although covering the
whole $3\sigma$ data, the $\theta_{23}$ are excluded at $2\sigma$
level when the $\delta$ lies in the rang $50^{\circ}\sim
80^{\circ}$($280^{\circ}\sim 310^{\circ}$). We also get
$0.015$eV$<m_{ee}<$ 0.040eV for
$0^{\circ}<\delta<80^{\circ}$($280^{\circ}<\delta<360^{\circ}$) and
$0.045$eV$<m_{ee}<$ 0.085eV for
$90^{\circ}<\delta<95^{\circ}$($265^{\circ}\sim 270^{\circ}$). No
strong bound on $J_{CP}$ are obtained. For the NH case, we have
$\delta$ constrained in the range $0^{\circ}\sim
40^{\circ}$($320^{\circ}\sim 360^{\circ}$) with
$\theta_{12}>34^{\circ}$, $|J_{CP}|<0.02$ and $90^{\circ}\sim
100^{\circ}$($260^{\circ}\sim 270^{\circ}$) with $|J_{CP}|\geq0.03$.
For both $\delta$ allowed region, the $\theta_{23}$ is below the
maximality and $0.005$eV$<m_{ee}<$ 0.080eV.

5)In Fig.\eqref{6CIH} and Fig.\eqref{6CNH}, we present the allowed
region for type-VI pattern. As the previous cases, the type-VI
pattern relates with the type-V pattern via $\mu-\tau$ symmetry and
therefore we have $\delta$ restricted to the range of
$85^{\circ}\sim 90^{\circ}$($100^{\circ}\sim 180^{\circ}$) and
$180^{\circ}\sim 260^{\circ}$($270^{\circ}\sim 275^{\circ}$). The
effective mass of $0\nu\beta\beta$ decay is the same as the ones of
type-V,i.e. $0.015$eV$<m_{ee}<$ 0.040eV. Just like the IH case, the
$\theta_{23}$ and $\delta$ of type-VI are qual to the
$\pi/4-\theta_{23}$ and $\delta+\pi$ of type-V. In this sense,
although the $\delta$ covers the $180^{\circ}$ which is at around
the best-fit value of CP-violating phase from the neutrino
oscillation experiments\cite{data}, the $\theta_{23}$ is up the
maximality and thus ruled out by $2\sigma$ results.

In Table.1, we present some generic predictions of all viable
textures at two value of $\delta$. i.e. $\delta\simeq\pi/2$ where
the Dirac CP symmetry is maximal violated and the best-fit point
$\delta\simeq \pi$. One can see from the Table that the value of
$\theta_{12}$, $\theta_{23}$ and $m_{ee}$ indicates important
phenomenological implication for the model selection. We are
particular interested in the type-III with the normal hierarchy and
the type-VI with the inverted hierarchy. In both cases, two
important feature emerges: (i) there leaves a space for
$\theta_{23}<\pi/4$ indicated by the experimental data at 2$\sigma$
level. (ii) the Dirac CP-violating phase $\delta$ is covered to its
best-fit value: $1.09\pi$\cite{data}. Despite (i) and (ii) are not
fully established yet, they are noteworthy in the model
building\cite{180} and deserve to be examined in future experiments.

In conclusion, the neutrino mass matrix with zero element and zero
cofactor are stable against the running of GREs and can be realized
by introducing discrete flavor symmetries with scalar
singlets\cite{minor2}. In this work, we carry out a numerical and
comprehensive analysis of the viable textures with the current
experimental data. We study the correlation between the Dirac
CP-violating phase $\delta$, three mixing angle and the
$0\nu\beta\beta$ effective mass $m_{ee}$. We examine the predictive
powers of these correlations in the future experiments and
demonstrate that they are essential in model selection. We present
some notable predictions for all the survived textures at
$\delta\simeq\pi/2$ and $\delta\simeq\pi$. Interestingly, the
type-III(NH) and type-VI(IH) are found to be phenomenologically
interesting by the fact that $\theta_{23}$ are possible located
below $\pi/4$ and the Dirac CP-violating phase $\delta$ covers its
best-fit value $1.09\pi$. We except that a cooperation between
theoretical study from the flavor symmetry point view and a
phenomenology study with updated experimental data will help us
reveal the structure of neutrino mass texture.

\begin{acknowledgments}
The author would like to thank S. Dev, R.R Gautam for the useful
discussion during this work.
\end{acknowledgments}


\begin{table}
\caption{\label{} Some generic predictions for the viable textures
at $\delta\approx\pi/2$ and $\delta\approx\pi$.}
\begin{tabular}{|c|c|c|}\hline
Texture & $\delta\approx \pi/2$ & $\delta\approx\pi$\\ \hline

type-I(IH)  & $0.044$eV$<m_{ee}<0.050$eV  & not allowed\\ \hline

type-II(IH) & $0.044$eV$<m_{ee}<0.050$eV  & not allowed\\ \hline

type-III(IH) & $\theta_{23}<\pi/4$, $0.050$eV$<m_{ee}<0.090$eV & not
allowed\\ \hline

type-III(NH) & $0.002$eV$<m_{ee}<0.030$eV & $\theta_{23}<\pi/4$,
$m_{ee}<0.007$eV\\ \hline

type-IV(IH) & $\theta_{23}>47^{\circ}$, $0.050$eV$<m_{ee}<0.090$eV &
$\theta_{12}>34^{\circ}$, $\theta_{23}>48^{\circ}$,
$0.015$eV$<m_{ee}<0.030$eV\\ \hline

type-IV(NH) & $0.002$eV$<m_{ee}<0.030$eV & $\theta_{23}>\pi/4$,
$0.007$eV$<m_{ee}<0.014$eV\\ \hline

type-V(IH) & $0.040$eV$<m_{ee}<0.080$eV & not allowed\\ \hline

type-V(NH) & $38^{\circ}<\theta_{23}<43^{\circ}$,
$0.025$eV$<m_{ee}<0.075$eV & not allowed\\ \hline

type-VI(IH) & $\theta_{23}>42^{\circ}$, $0.050$eV$<m_{ee}<0.080$eV &
$\theta_{23}>40^{\circ}$, $0.010$eV$<m_{ee}<0.035$eV\\ \hline

type-VI(NH) & $\theta_{23}>47^{\circ}$, $0.030$eV$<m_{ee}<0.075$eV &
$\theta_{12}>34^{\circ}$, $\theta_{23}>49^{\circ}$,
$0.010$eV$<m_{ee}<0.025$eV\\ \hline
\end{tabular}
\end{table}

\begin{thebibliography}{21}
\bibitem[1]{neu1}
Q.R. Ahmad \textsl{et al.}(SNO Collaboration), Phys. Rev. Lett {\bf
89}, 011301(2002); K. Eguchi \textsl{et al.} (KamLAND
Collaboration), Phys. Rev. Lett {\bf 90}, 021802(2003); M.H. Ahn
\textsl{et al.} (K2K Collaboration), Phys. Rev. Lett {\bf 90},
041801(2003).

\bibitem[2]{neu2}
F.P. An \textsl{et al.} (DAYA-BAY Xollaboration), Phys. Rev. Lett.
{\bf 108}, 171803(2012).

\bibitem[3]{neu3}
J.K. Ahn \textsl{et al.} (RENO Collaboration), Phys. Rev. {\bf
D108}, 191802(2012).

\bibitem[4]{WMAP}
G. Hinshaw \textsl{et al.} (WMAP Collaboration), arXiv: 1212.5226.

\bibitem[5]{Planck}
P.A.R. Ade \textsl{et al.} (Planck Collaboration), arXiv: 1303.5076.

\bibitem[6]{NDD}
S.M. Bilenky and C. Giunti, Mod. Phys, Lett. {\bf A16},
1230015(2012).

\bibitem[7]{seesaw}
H. Fritzsch, M. Gell-Mann, and P. Minkowski, Phys. Lett. {\bf B59},
256(1975); P. Minkowski, Phys. Lett. {\bf B67}, 421(1977); T.
Yanagida, in \textsl{Proceedings of Workshop on Unified Theory and
the Baryon Number of the Universe}, edited by O. Sawada and A.
Sugamoto(KEK, Tsukuba, 1979), p. 95; M. Gell-Mann, P. Ramond, and
Slansky, in \textsl{Supergravity}, edited by P. van. Nieuwenhuizen
and D.Z. Freeman (North-Holland, Amsterdam,1979), p. 315; R.N.
Mohapatra and G. Senjanovic, Phys, Rev. Lett. {\bf 44}, 912(1980);
J. Schechter and J. W. F. Valle,  Phys. Rev. {\bf D22}, 2227(1980);
J. Schechter and J. W. F. Valle,  Phys. Rev. {\bf D25}, 774(1982).

\bibitem[8]{zero}
P.H. Frampton, S. L. Glashow, and D. Marfatia, Phys. Lett. {\bf
B536}, 79(2002); Z,-z. Xing, Phys. Lett. {\bf B530}, 159(2002); M.
Randhawa, G. Ahuja, and M. Gupta, Phys. Lett. {\bf B643}, 175(2006);
A. Merle, and W. Rodejohann, Phys. Rev. {\bf D73}, 073012(2006); S.
Dev, S. Kumar, S. Verma, and S. Gupta, Phys. Rev. {\bf D76},
013002(2007); S. Dev, S. Kumar, S. Verma, and S. Gupta, Nucl. Phys.
{\bf B784}, 103(2007); G. Ahuja, S. Kumar, M. Randhawa, M. Gupta,
and S. Dev, Phys. Rev. {\bf D76}, 013006(2007); S. Dev, S. Kumar,
Mod. Phys, Lett. {\bf A22}, 1401(2007); S. Kumar, Phys. Rev. {\bf
D84}, 077301(2011); P.O. Ludl, S. Morisi, and E. Peinado, Nucl.
Phys. {\bf B857}, 411(2012); W. Grimus. and P.O. Ludl,
arXiv:1208.4515; D. Meloni, and G. Blankenburg, Nucl. Phys. {\bf
B867}, 749(2013); H. Fritzsch, Z.-z. Xing, and S. Zhou, J. High
Energy Phys. 09 (2011)083.

\bibitem[9]{hybrid}
S. Kaneko, H. Sawanaka, and M. Tanimoto, J. High Energy Phys. 08
(2005)073; S. Dev, S. Verma, and S. Gupta, Phys. Lett. {\bf B687},
53(2010); S. Dev, S. Gupta, and R.R. Gautam, Phys. Rev. {\bf D82},
073015(2010); S. Goswami, S. Khan, and A. Watanable, Phys. Lett.
{\bf B687}, 53(2010), W. Grimus, and P. O. Ludl, arXiv: 1208.4515.

\bibitem[10]{hybrid2}
J.-Y. Liu and S. Zhou, Phys. Rev. {\bf D87}, 093010(2013).


\bibitem[11]{sum}
X.-G. He and A. Zee, Phys. Rev. {\bf D68}, 037302(2003).

\bibitem[12]{det}
G.C. Branco, R. Gonzalez Felipe, F.R. Joaquim, and T. Yanagida,
Phys. Lett. {\bf B562}, 265(2003); B.C. Chauhan, J. Pulido, and M.
Picariello, Phys. Rev. {\bf D73}, 053003(2006).

\bibitem[13]{minor}
L. Lavoura, Phys. Lett. {\bf B609}, 317(2005); E.I. Lashin and N.
Chamoun, Phys. Rev. {\bf D78}, 073002(2008); E.I. Lashin and N.
Chamoun, Phys. Rev. {\bf D80}, 093004(2009); S. Dev, S. Gupta, and
R.R. Gautam, Mod. Phys, Lett. {\bf A26}, 501(2011); S. Dev, S.
Gupta, R.R. Gautam, and L. Singh, Phys. Lett. {\bf B706}, 168(2011);
T. Araki, J. Heeck, and J. Kubo, J. High Energy Phys. 07 (2012)083;
S. Verma, Nucl. Phys. {\bf B854}, 340(2012); S. Dev, R.R. Gautam,
and L. Singh, arXiv: 1309.4219;

\bibitem[14]{minor2}
S. Dev, S. Verma, S. Gupta, and R.R. Gautam, Phys. Rev. {\bf D81},
053010(2010); J. Liao,D. Marfatia, K. Whisnant, arXiv: 1311.2639.

\bibitem[15]{tra}
H.A. Alhendi, E.I. Lashin, and A.A. Mudlej, Phys. Rev. {\bf D77},
013009(2008).

\bibitem[16]{co}
S. Dev, R.R. Gautam, and L. Singh, Phys. Rev. {\bf D87},
073011(2013).

\bibitem[17]{hyco}
S. Dev, R.R. Gautam, and L. Singh, Phys. Rev. {\bf D88},
033008(2013); W. Wang, Eur. Phys. J. {\bf C73}, 2551(2013).

\bibitem[18]{PMNS}
B. Pontecorvo, Zh. Eksp. Teor. Fiz. {\bf 33}, 549(1957); Z. Maki, M.
Nakagawa, and N. Sakata, Prog. Theor. Phys. {\bf 28}, 870(1962).

\bibitem[19]{Jas}
C. Jarlskog,  Phys, Rev. Lett. {\bf 55}, 1039(1985).

\bibitem[22]{data}
G.L. Fogli, E. Lisi, A. Marrone, D. Montanino, A. Palazzo, and A.M.
Rotunno, Phys. Rev. {\bf D86}, 013012(2012); D. V. Forero, M.
Tortola, and J. W. F. Valle, Phys. Rev. {\bf D86}, 073012(2012); M.
C. Gonzalez-Garcia, Michele Maltoni, Jordi Salvado, and Thomas
Schwetz, J. High Energy Phys. 12 (2012)123.


\bibitem[23]{HM}
H.V. Klapdor-Kleingrothaus, A. Dietz, H.L. Harney, and I.V.
Krivosheina, Mod. Phys. Lett. {\bf A16}, 2409(2001).

\bibitem[24]{NND2}
C.E. Aalseth \textsl{et al}. Mod. Phys. Lett. {\bf A17}, 1475(2002);
F. Feruglio, A. Strumia, and F. Vissani, Nucl. Phys. {\bf B637},
345(2002).

\bibitem[25]{mut}
T. Fukuyama and H. Nishiura, arXiv: 9702253; R. N. Mohapatra and S.
Nussinov, Phys. Rev. D60, 013002, (1999); E. Ma and M. Raidal, Phys.
Rev. Lett. 87, 011802 (2001); C. S. Lam, Phys. Lett. B507, 214
(2001); K. R. S. Balaji, W. Grimus and T. Schwetz, Phys. Lett. B508,
301 (2001); W. Grimus and L. Lavoura, Acta Phys. Pol. B32, 3719
(2001).

\bibitem[26]{180}
L. Lavoura, W. Rodejohann, A. Watanabe, Phys. Lett. {\bf B726},
352(2013).

\end{thebibliography}
\end{document}